\documentstyle[proceedings,psfig]{gamcrckapb}
\font\smallital=cmti9
\begin{opening}
\title{Preliminary galaxy extraction \protect\\ from DENIS images}
\author{G. A. MAMON}
\author{V. BANCHET}
\author{M. TRICOTTET}
\author{D. KATZ}
\institute{Institut d'Astrophysique\\
98 bis Bd Arago, F-75014 Paris, France}
\end{opening}
\runningtitle{Preliminary galaxy extraction from DENIS}

\begin{document}
\begin{abstract}
The extragalactic applications of NIR surveys are summarized with a focus on
the ability to map the interstellar extinction of our Galaxy.
Very preliminary extraction of galaxies on a set of 180 consecutive images is
presented, and the results illustrate some of the pitfalls in attempting an
homogeneous extraction of galaxies from these wide-angle and shallow surveys.
\end{abstract}

\section{Introduction}

Near Infrared (NIR) surveys such as DENIS and 2MASS are extremely useful for
extragalactic 
astronomy and cosmology, for mainly two reasons.
\begin{enumerate}
\item The extinction in the NIR bands is 11\% ($K_s$) to 28\% ($J$) of the
visual extinction.
This allows one with a wider view of the Universe, then allowed by optical
surveys, limited to high galactic latitudes because of the extinction from
interstellar dust of our Galaxy.
This also allows a better view of external galaxies, as the light is
unhampered by extinction from their own interstellar dust.
\item While optical light (in particular blue and ultraviolet) is extremely
sensitive to massive very young (million year old) stars, NIR light is much
less sensitive to recent star formation.
\end{enumerate}
A caveat is in order for this second point: 
NIR light turns out to be more sensitive to 10 million year old bursts of
star formation than other wavebands, because the very massive young stars
evolve in that timescale onto the giant branch.
This is well illustrated in synthetic spectra of iso-eval stellar
populations (e.g. Bruzual \& Charlot, 1993, Fig. 4a), in which the
dispersion in 
waveband averaged intensity ({\it i.e.\/,} broad-band flux) for different
ages of the stellar population is smallest in the NIR, with the exception of
the epoch at 10 million years.
See also Knapen (in these proceedings) for an illustration of NIR luminous
regions with recent star formation in the nuclei of external galaxies.

One could go to the mid-IR to avoid extinction by dust even further, but
mid-IR light is sensitive to thermal emission from warm dust associated with
recently formed stars and starting around the $K$ band at 2.2 microns,
ground-based observations are limited by thermal emission from the instrument.

\section{Extragalactic applications}

The applications of NIR surveys for extragalactic astronomy and cosmology
have been described elsewhere (Mamon 1994, 1995, 1996; Schneider in these
proceedings) and
are briefly outlined again here.

The main extragalactic applications foreseen with DENIS are

\begin{itemize}

\item {\sl A statistical sample of properties of NIR galaxies\/}

The large sample size will help study correlations between properties.

\item {\sl The 2D structure of the local Universe\/}.

Catalogs of groups and clusters will be obtained from the galaxy lists
extracted from the survey images.
Statistical measures of large-scale structure will be obtained (such as the
angular correlation function and higher order functions, counts in cells, and
topological measures).
It will be interesting to see how all this 2D structure will vary with
waveband, which will hopefully tell us about how structure is a function of
waveband and indicate possible biases when going from one waveband to
another.
The alternative is that any such difference in 2D structure may be a
reflection of selection effects, but we are working hard on avoiding this.

\item {\sl Color segregation\/}

Instead of studying structure versus waveband, one can study the inverse
problem of understanding colors as a function of structure, hence
environment.
Color segregation is a potentially powerful probe of three dimensional
morphological segregation of galaxies in the Universe ({\it e.g.\/,} the
cores of clusters being richer in ellipticals and poorer in spirals).
Monte-Carlo tests are planned to find out 
whether one can recover such a morphological segregation
from the 2D color segregation information that will be obtained from DENIS
(despite the loss of information from projection effects, extinction,
k-corrections, and tidally triggered star formation).
If yes, then using $I-J$ versus environment, with a sample of over $10^5$
galaxies, DENIS would provide the largest sample for studying morphological
segregation (typically limited to $10^{3.5}$ galaxies today).

\item {\sl Normalization of galaxy counts at the bright-end}

Counting galaxies as a function of apparent magnitude provides better results
at the faint-end than at the bright-end, simply because the bright-end
suffers from very poor statistics (in the local uniform Universe, galaxy
counts rise roughly as ${\rm dex} [0.6\,m]$).
If our Local Group sits in an underdense region, we should see a lack of
galaxies at the very bright end of the galaxy counts, which is brighter than
the DENIS complete/reliable extraction limit.
The accepted standards for galaxy counts, arising from the APM (Maddox et
al. 1990) and COSMOS (Heydon-Dumbleton et al. 1989) surveys have yielded
bright-end counts that were 
inconsistent with the intermediate magnitude counts, for a uniform Universe
with no abnormally strong galaxy evolution.
In recent studies (Bertin \& Dennefeld 1996; Gardner et
al. 1996), the bright-end normalization in $B$ is roughly twice as large as
in the 
APM and COSMOS counts (the discrepancy is apparently due to poor correction
for non-linear plate response in the first two studies), and in agreement
with simply extrapolated fainter counts.
However, the error bars still remain large, and very wide-angle surveys such
as DENIS and 2MASS will bring them down.
Moreover, two recent studies (Lidman \& Peterson 1996; Gardner et al. 1996)
display strong discrepancy of the bright-end 
counts in $I$ (see. Figure \ref{counts}).
\begin{figure}[ht]
\psfig{file=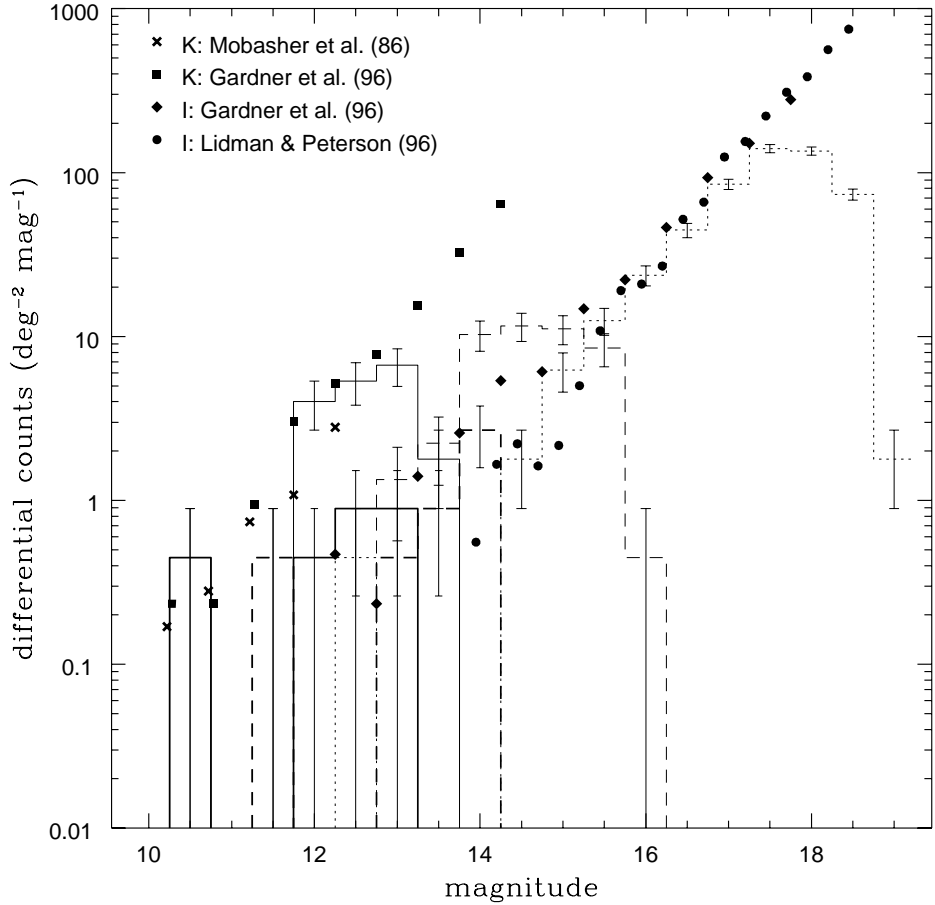,width=13cm}
\caption{Counts of candidate galaxies for strip 3081 of 180 consecutive
images. 
{\it Solid\/}, {\it dashed\/}, and {\it dotted\/} {\it thin\/} histograms
show our DENIS 
counts in $K$, $J$, and $I$.
The {\it thick\/} histograms refer to objects also seen in the bluer band(s)
($I$ and $J$ for $K$ and $I$ for $J$).
For comparison are also shown $K$ counts by Mobasher et al. (1986)
and Gardner et al. (1996), and $I$ counts by Lidman \& Peterson (1996) and
Gardner et al. (1996). All surveys have similar areas, so the error bars have
roughly the same sizes for given ordinate.
}
\label{counts}
\end{figure}

\item {\sl Mapping interstellar extinction}

Among the many ways one can map the interstellar extinction in our Galaxy,
one is to use galaxy counts (Burstein \& Heiles 1978, 1982), since the count
normalization is shifted downwards when galaxies are
extinguished. 
Galaxy counts are tricky extinction calibrators, because of intrinsic
variations of galaxy counts (the correlation of the galaxy distribution
implies fluctuations of counts in cells that are greater than Poisson).
Also, the low sensitivity of NIR surveys implies the need of very large
solid angles for significant counts, hence any extinction map would be of
very poor angular resolution.
On the other hand, galaxy colors may provide the best way to map the
interstellar extinction, as they redden with extinction (this method is
hampered by any intrinsic color segregation).
The maximum extinction that can be probed using $I-J$ colors is
$$
A_V^{\rm max} = {I_{\rm lim}-J_{\rm lim} - (I-J)_0\over E(I-J)/A_V} \simeq
9
\ ,
$$
with
$I_{\rm lim} \simeq 17.0$,
$J_{\rm lim} \simeq 14.5$,
$(I-J)_0 \simeq 0.7$,
and
$E(I-J)/A_V = 0.2$.
Beyond this extinction, we start missing galaxies in the more extinguished
$I$ band. 
The uncertainty on the extinction comes from the uncertainty on the mean
galaxy color, which in turn arises from the width of the color distribution
and the photometric errors.
With $\Delta I \simeq \Delta J \simeq 0.2$, and a similar dispersion in
intrinsic colors, a sample of $N$ galaxies will
yield an uncertainty
\begin{equation}
\Delta A_V = N^{-1/2} {\Delta (I-J)\over  E(I-J)/A_V} \simeq 2\,N^{-1/2} \ .
\end{equation}
Given the expected unextinguished counts in the $J$ band, 
$N(<J) \simeq 5 \times {\rm dex} [0.6(J-14.5)]\,\rm deg^{-2}$,
using $A_J = 0.28\,A_V$, the extinguished counts fall by a factor
${\rm dex}(0.6\times 0.28 A_V)$, while the galaxy extraction limit will fall
by roughly $0.28\,A_V$ magnitude, hence the counts of extracted galaxies in
extinguished regions will vary roughly as 
\begin{equation}
N \simeq 5 \,{\rm dex} \left (1.2\times 0.28 A_V \right ) \,\rm
(\Delta\theta)^2  ,
\end{equation}
where $(\Delta \theta)^2$ is the solid angle in $\rm deg^2$.
For an accuracy $\Delta A_V$, one requires (eqs. [1] and [2]) an angular
resolution
\begin{equation}
\Delta \theta \simeq {0.9\over \Delta A_V} {\rm dex}\left [-\left (0.6\times
0.28 A_V \right ) \right ] \ {\rm deg} \ .
\end{equation}
The results are shown in Table 1.
\end{itemize}
\begin{table}[ht]
\caption{Angular resolution required for mapping extinction (eq. [3])}
\begin{center}
\begin{tabular}{l c c c c c c c c}
\hline
$A_V$ & 1 & 2 & 5 & 5 & 5 & 9 & 9 & 9 \\
$\Delta A_V$ & 0.2 & 0.2 & 0.1 & 0.2 & 0.5 & 0.1 & 0.2 & 0.5 \\
$\Delta \theta$ (deg) & 3.03 & 2.06 & 1.29 & 0.65 & 0.26 & 0.28 & 0.14 & 0.06
\\ 
\hline
\end{tabular}
\end{center}
\end{table}

\section{Galaxy pipelines}

\subsection{Goal}

The DENIS galaxy pipeline is still under development.
It is a 2-pass pipeline described below:

In the first pass, all sources extracted in the LDAC
pipeline are checked for a minimum area above threshold, required for
reliable galaxy detection (the values are obtained by image simulations).
Kron photometry is applied on these sources with parameters optimized from
image simulations.
Star galaxy separation is performed in two manners:
1) by classical methods, using a combination of maximum pixel-intensity
(corrected for the position of the object centroid relative to the pixel
center), area above threshold, and full-width half-maximum, given the
magnitude.
2) by neural networks trained on simulated images (Bertin 1996).
So far, neural networks prove superior, but also make mistakes for bright or
intermediate flux galaxies that the
classical methods avoid. 

In the second pass, stars are removed from the images, by masking the
saturated stars and subtracting the PSF from non-saturated ones.
Then large-scale smoothing is applied to the images, which allows one to
recover face-on late-type spiral galaxies, that are otherwise invisible in
$K$.
Object detection is done with parameters optimized to avoid fragmentation of
the brighter objects (due to photon noise).
Kron photometry is performed again.
Obviously no star galaxy separation is required for this second list,
although checks will be made.

\subsection{Preliminary pipeline}
\label{sec:prelimpipe}
We now present preliminary results from the extraction with a preliminary
pipeline, which is as follows:

We first filter the images to remove cosmics and bad pixels.
We then use the easy-to-use {\sl SExtractor\/} package (Bertin \& Arnouts
1996). 
Again we set a minimum area above threshold for galaxies and smooth the
images with a conical $3\times3$ filter in the $I$ band, and the convolution
of that filter with a $3\times3$ boxcar in the $J$ and $K$ bands (because
a $3\times3$ boxcar is required to transform the interlaced images of 9
sub-images in these 2
bands into images obtained by co-adding the sub-images).
The threshold is set at $3\,\sigma$ on the smoothed images (as optimized from
simulated images), and Kron photometry is applied with parameters optimized
from the simulated images.
{\sl SExtractor\/} uses neural networks for the star galaxy separation, and
assumes a constant point-spread-function (PSF) within and among strips of 180
consecutive images.
The results below are for objects with {\sl SExtractor\/} stellarity index
$<0.1$. 

\section{Results on two strips}

\begin{figure}[ht]
\psfig{file=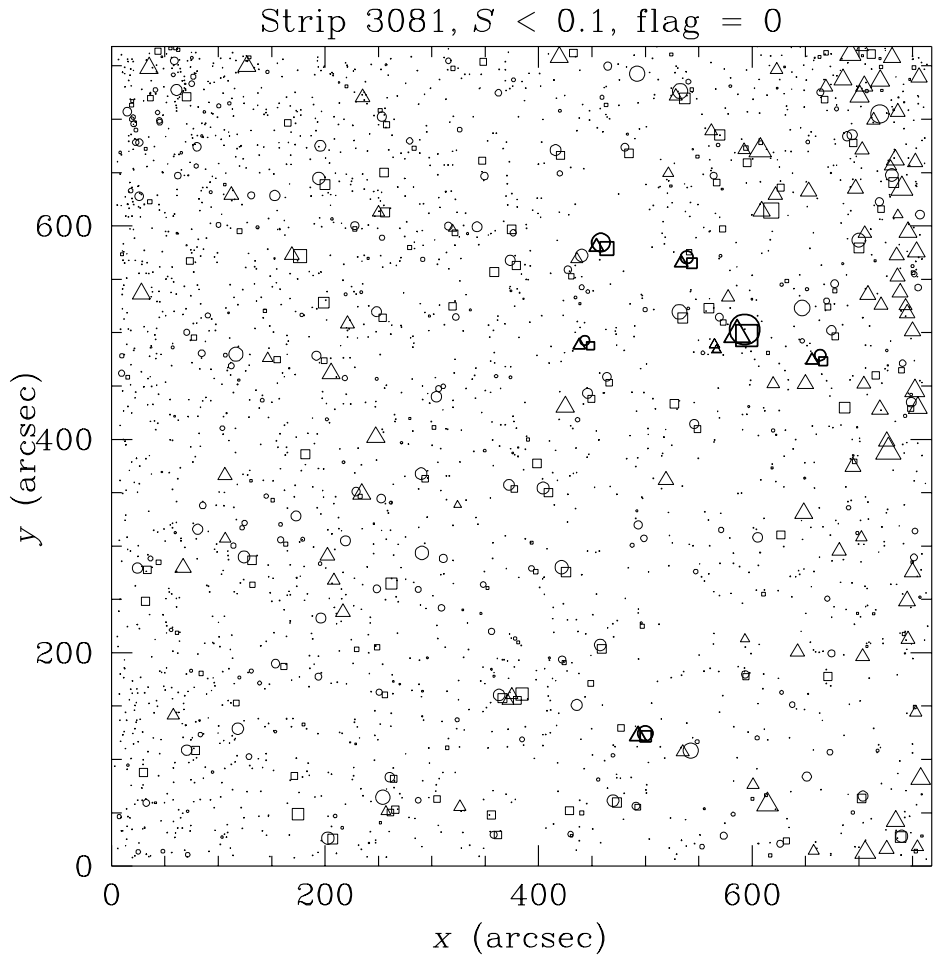,width=13cm}
\caption{Positions of galaxy candidates in 180 superposed images.
Objects extracted in $I$, $J$, and $K$ are shown as {\it circles\/}, {\it
squares\/}, and {\it triangles\/}, respectively. The sizes of the symbols are
scaled to their magnitude, with the same normalization for the brightest
object of each waveband. Objects that appear galaxy-like in all three
wavebands are shown with {\it thick symbols\/}.}
\label{positions}
\end{figure}

Figure \ref{positions} shows the superposition of the positions of candidate
galaxies  
over 180 consecutive images of a good quality high galactic latitude strip
(3081).
Careful inspection of Figure \ref{positions} indicates that there is
vignetting in $I$ and 
an excess of objects on the right-side of the frames in $K$, which is due to
a small linear variation of the PSF along Right Ascension in this strip.

\begin{figure}[ht]
\begin{center}
\psfig{file=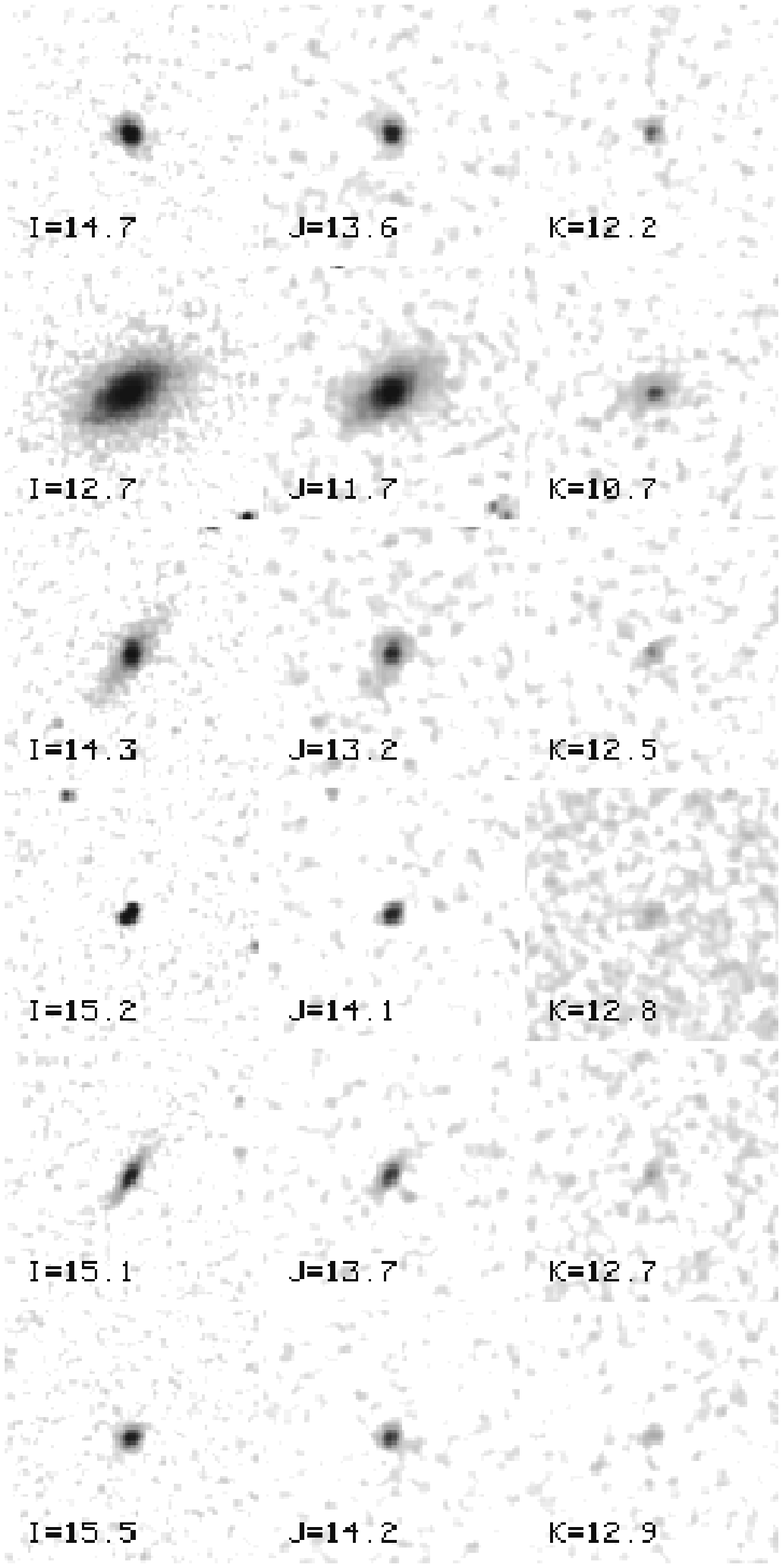,height=15cm}
\end{center}
\caption{Six candidate galaxies from strip 3081 of 180 consecutive images in
$I$ (left), $J$ (middle), and $K$ (right).
Kron photometry is given for each object.
Greyscale is logarithmic from 1.5 to 50 $\sigma$.
Images are smoothed with the filters discussed in section
\ref{sec:prelimpipe}.
}
\label{stamps}
\end{figure}

Six candidate galaxies are seen in all three wavebands.
Five out of six are in the same upper-right quadrant of their respective
images, which is probably a signature of PSF variations.
The six candidates are shown in Figure \ref{stamps}.
The $K$ images clearly limit the galaxies to their central regions.
The fourth candidate galaxy from the top appears to be a double star (see the
$I$ image on the left).

\begin{figure}[ht]
\begin{center}
\psfig{file=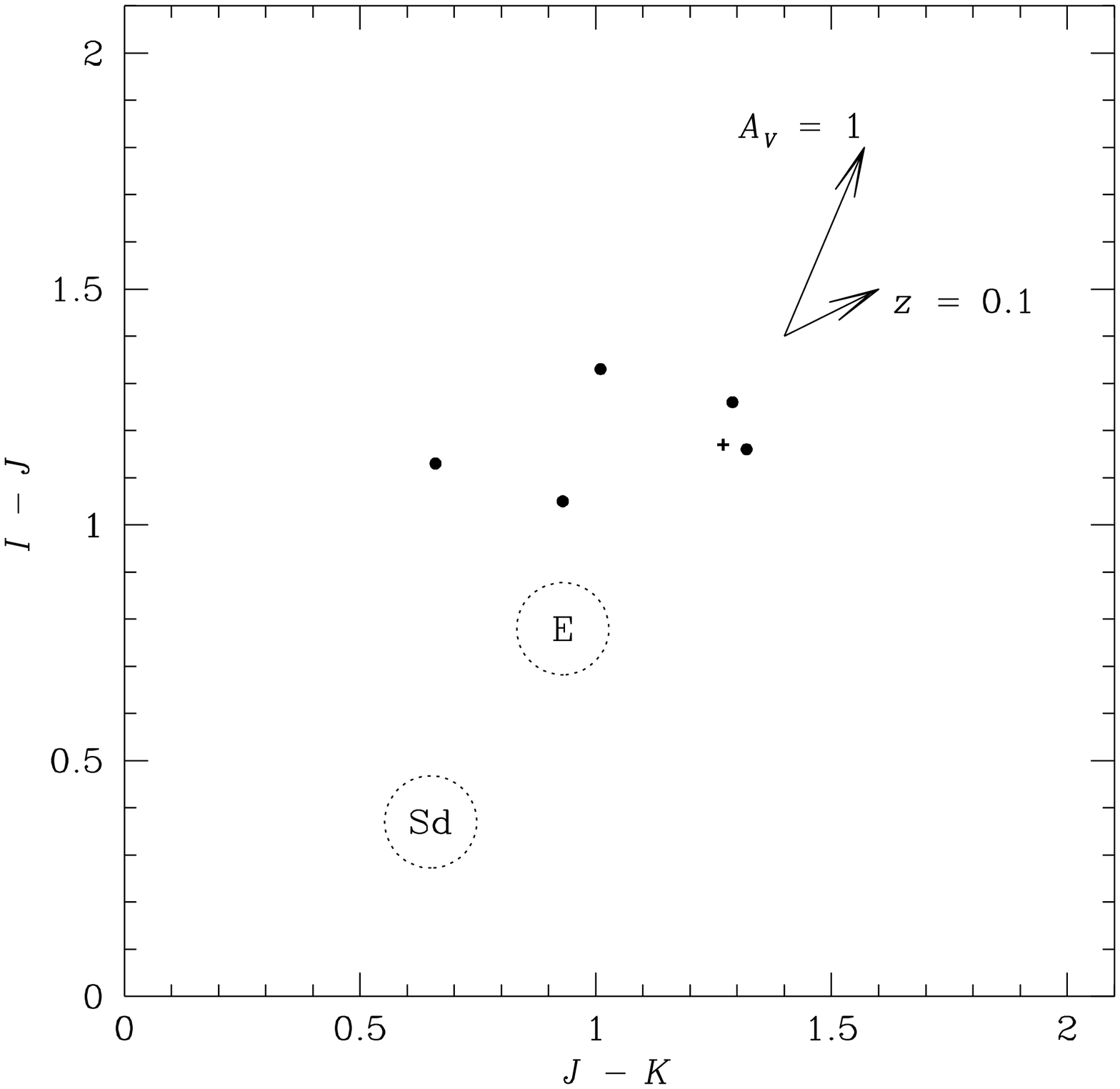,height=9cm}
\end{center}
\caption{Color-color diagram of strip 3081.
The {\it cross\/} is for the candidate galaxy appearing as a double star.
The typical colors (from the compilation of Yoshii \& Takahara 1988, see also
Peletier in these 
proceedings) of ellipticals (E) and late-type
spirals (Sd) are shown. 
Arrows indicate one visual-magnitude of extinction ($A_V=1$) and the effects
of differential $k$-correction ($z=0.1$).
}
\label{colorcolor}
\end{figure}

The colors of the six galaxy candidates are shown in Figure \ref{colorcolor}.
While at least two of the galaxies in Figure \ref{stamps} are clearly edge-on
spirals, their colors are slightly redder than ellipticals.
Because the low sensitivity in $K$ limits the visibility and photometry of
galaxies to their central regions, one would have expected that the $J-K$
colors be too blue, which is not seen.

In Figure \ref{counts} are shown galaxy counts for this strip (we omit here
objects at 
less than $100''$ from the image edges, yielding a solid angle of $4.5\,\rm
deg^2$ for the strip). 
The `raw' $K$ counts agree better with previous work than the `constrained'
$K$ counts.
These raw counts are however likely to be severely contaminated by stars.
The constrained object list is thus highly incomplete.
The $I$ band counts show a good agreement with those of Lidman \& Peterson
(1996) and Gardner et al. (1996), suggesting a completeness limit of $I \simeq
16.75$.

\section{Discussion and conclusions}

The incompleteness of our extraction in $K$ illustrates the inadequacy of our
preliminary extraction  pipeline.
We processed a second strip, of poor quality (with strong PSF variations
across the images) and obtained `constrained' counts in $K$ that matched
better previous determinations, and suggested a completeness limit of $K
\simeq 12.25$.
We have recently begun to incorporate modeling of the PSF (see Borsenberger,
in these proceedings) over a number of
subsequent images into our star/galaxy separators.
Since over 99\% of objects are stars, we perform an iterative asymmetric
rejection of large PSFs before modeling it.
Preliminary tests yield rms residuals less than $0.15''$, after optimizing
for the minimum area of objects to consider and the number of
subsequent images on which the modeling is done.
We expect the new star/galaxy separators to have much more uniform galaxy
distributions over the images in comparison with what is shown in Figure
\ref{positions}.

We thank Emmanuel Bertin for numerous useful discussions.

\end{document}